\documentclass[fleqn,10pt]{wlscirep}
\usepackage[utf8]{inputenc}
\usepackage[T1]{fontenc}

\usepackage{caption}
\usepackage{subcaption}

\newcommand{\mvec}[1]{\bm{#1}}
\newcommand{\iu}[0]{\mathrm{i}}
\newcommand*\diff{\mathop{}\!\mathrm{d}}
\usepackage{bm}

\title{Polariton spectroscopy at the diamond K-edge via X-ray parametric down-conversion}

\author[1,2,+]{Fridtjof Kerker}
\author[3,+,*]{Dietrich Krebs}
\author[2]{Xenia Brockm\"uller}
\author[1,3]{Ankita Negi}
\author[4]{Christoph J. Sahle}
\author[4]{Blanka Detlefs}
\author[1,2,3,*]{Christina B\"omer}
\affil[1]{The Hamburg Centre for Ultrafast Imaging, 22761 Hamburg, Germany}
\affil[2]{Department of Physics, University of Hamburg, 22607 Hamburg, Germany}
\affil[3]{Deutsches Elektronen-Synchrotron DESY, 22603 Hamburg, Germany}
\affil[4]{ESRF, The European Synchrotron, 71 Avenue des Martyrs, CS40220, 38043 Grenoble Cedex 9, France}

\affil[+]{These authors contributed equally to this work}
\affil[*]{dietrich.krebs@desy.de \\ christina.boemer@desy.de}

\begin{abstract}
It has recently been shown that x-ray parametric down-conversion (XPDC) provides access to high-energy polaritons, resulting from the hybridization of down-converted photons with electronic excitations in a nonlinear medium. Here, we present a spectrally resolved study of this effect around the K-shell absorption edge in diamond. Our results exhibit pronounced signatures of polaritonic hybridization, which we visualize by introducing a polariton spectral map and analyze by help of theoretical modelling. We find that the hybridization at this absorption edge results in substantially higher coupling strength than previously reported for a non-resonant case and reaches well into the strong-coupling regime. In addition, we demonstrate how our measurements of polaritonic XPDC allow us to extract the refractive index for bulk diamond at high spectral resolution around the carbon K-edge.

\end{abstract}
\begin{document}

\flushbottom
\maketitle

\thispagestyle{empty}

\section*{Introduction}

Nonlinear optics has given rise to a plethora of processes for up- and down-conversion of light \cite{bloembergen_nonlinear_1965,franken_generation_1961,1967ZhPmR...6..575A,harris_observation_1967,akhmanov_quantum,magde_study_1967,giordmaine_tunable_1965,akhmanov1965observation}, which have - in turn - found broad application across science and technology \cite{menzel_photonics_2007,li_down-converted_2025,wegner_nif_2004,zhang_recent_2020,dudley2024nonlinear}. Among these processes, spontaneous parametric down-conversion (SPDC) holds a special place given its fundamentally quantum nature and the unique access to non-classical states of light and matter that it enables:
First of all, SPDC provides a source of entangled photon pairs \cite{kwiat_new_1995,rubin_theory_1994}, ubiquitously used in quantum technology today \cite{valencia_large-scale_2025,lu_counter-propagating_2025,chakraborty_towards_2025,lyu_tunable_2025} - yet in addition, it also provides a gateway to the quantum hybridization of light with matter resulting in polaritons \cite{hopfield_theory_1958,basov_polariton_2020}.
The intimate connection of parametric down-conversion and light-scattering by polaritons was originally identified by Klyshko at optical wavelengths \cite{klyshko_scattering_nodate} and recently transferred to the EUV-regime using x-ray parametric down-conversion (XPDC) \cite{krebs_x-ray_2025}.
This development allows for unprecedented studies of strong-coupling phenomena in the EUV to soft x-ray range, where other methods to this end were entirely missing previously. 
Here, we extend this work further and demonstrate that the analogy with optical polaritons runs deeper still - even allowing for the transfer of techniques like polariton k-spectroscopy \cite{coffinet_coherent_1969,kulevsky_light_1975,aktsipetrov_frequency-angle_nodate,chekhova_study_1993} from optical to x-ray wavelengths. Following a brief introduction to our experimental setup for momentum-resolved detection of XPDC, 
we present down-conversion measurements at the carbon K-edge in diamond. 
Visualizing these in a two-dimensional polariton spectral map - inspired by optical k-spectroscopy - we find rich features that we address with subsequent theoretical modelling. 
Finally, we use our model to gain new insights on strong coupling in the soft x-ray regime and demonstrate how it can be used to extract the refractive index of otherwise inaccessible bulk materials. 

\section*{Results}
\subsection*{Accessing polaritons via XPDC}
\begin{figure}[ht]
\centering
\includegraphics[width=0.9\linewidth]{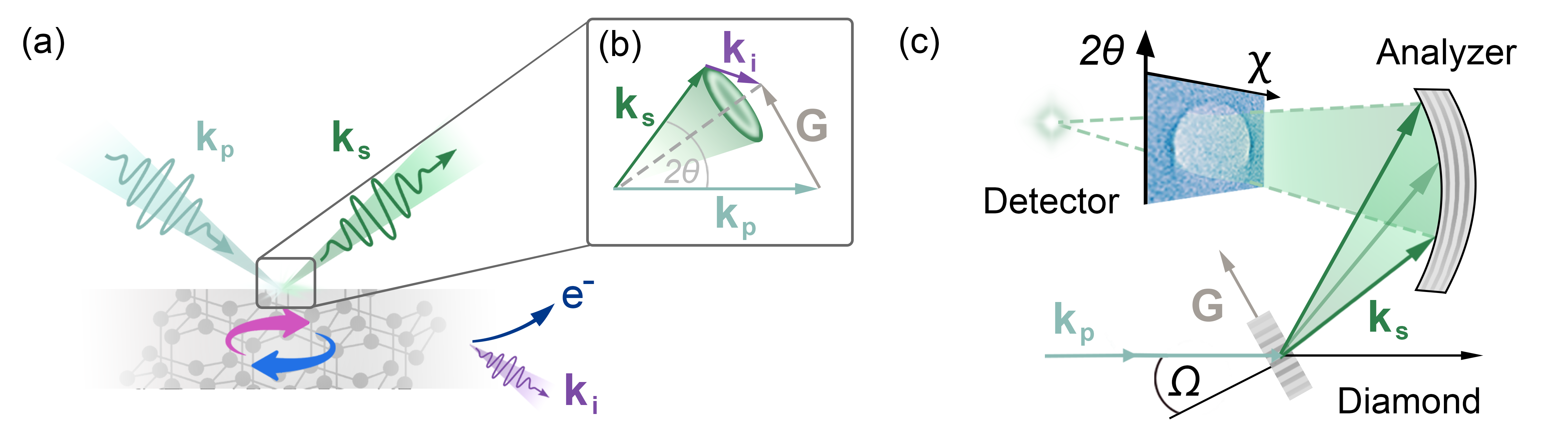}
\caption{Emergence and detection mechanism of polaritons via XPDC: (a) The pump photon ($\mvec{k_p}$) spontaneously splits into a correlated photon pair of lower energies, the signal ($\mvec{k_s}$) and idler ($\mvec{k_i}$). Subsequent absorption and re-emission of the EUV-idler quantum within the material forms the polariton - a hybridization of photonic and electronically excited states. The higher-energy signal photon is diffracted and carries imprints of this polariton. 
(b) Phase matching condition for XPDC, with the pump photon ($\mvec{k_p}$) diffracting nonlinearly off the reciprocal lattice vector ($\mvec{G}$) and splitting into the signal ($\mvec{k_{s}}$) and the idler ($\mvec{k_i}$). Rotation symmetry causes emission into cones (green for the signal, purple for the idler). (c) The experimental realization of XPDC has the pump photon impinging on the diamond sample under an angle $\Omega$ with respect to the selected lattice planes.
The generated signal cone ($\mvec{k_s}$, green) is reflected by the analyzer, filtering at \mbox{$\hbar \omega_s = 9.69$~keV} while suppressing scattering processes at other energies. The analyzer’s spherical shape preserves the momenta of the signal cone and images it onto a 2D detector positioned out-of-focus, thus resolving the scattering signature along both axis $2\theta$ and $\chi$.}
\label{fig:setup}
\end{figure}

In order to gain (spectroscopic) access to polaritons at EUV and soft x-ray wavelengths, we use the process of x-ray parametric down-conversion (XPDC). 
This nonlinear scattering process provides a unique mechanism to both launch and probe high-energy polaritons \cite{krebs_x-ray_2025}, which is schematically illustrated across Fig.~\ref{fig:setup}. During XPDC, an x-ray pump photon ($\mvec{k_p}$) spontaneously splits into a highly-correlated pair of lower-energy photons - the signal ($\mvec{k_s}$) and the idler ($\mvec{k_i}$). For highly asymmetric splitting ratios, the idler photon’s energy ($\hbar \omega_i$) can be within the soft x-ray regime, where it is readily absorbed by the surrounding sample.
  
Subsequent re-emission and -absorption will lead the idler photon to hybridize with electronic excitations of its environment, forming a polaritonic state (illustrated by circular arrows in Fig.~\ref{fig:setup} a). While this idler polariton is ultimately lost to detection, signatures of its hybridization remain imprinted onto the scattered signal photon, owing to the strong correlation of the XPDC photon-pair \cite{krebs_x-ray_2025}. These imprints are particularly pronounced, if the idler energy is tuned to electronic resonances or excitation edges, where the associated polaritonic signatures exhibit variegated features that we shall explore in detail for the carbon K-edge in diamond below.

For the XPDC process to occur, energy has to be conserved among the produced photon pair $\omega_p \Rightarrow \omega_s + \omega_i$. Together with momentum conservation $\mvec{k_p} + \mvec{G} = \mvec{k_s} + \mvec{k_i}$, this constitutes the so-called phase-matching condition (cf. Fig.~\ref{fig:setup} b).
Here, $\mvec{G}$ denotes a reciprocal lattice vector of the crystalline sample, on which the nonlinear scattering takes place. Furthermore, all photon energies are connected to their pertaining wave vectors via dispersion relations of the usual form $\omega_j = \hbar c \,  | \mvec{k_j} | / n_j$, wherein the refractive index is trivial for the hard x-ray photons ($n_p \approx 1 \approx n_s$) but turns out to be influential for the soft x-ray idler photon. Combining the above conditions allows for signal and idler photons to be produced on opposing cones, as energy conservation only fixes their magnitudes $| \mvec{k_s} |$ and $| \mvec{k_i} |$ but leaves a rotational degree of freedom for the emitted wave vectors around the joint vector $\mvec{k_s} + \mvec{k_i}$ (Fig.~\ref{fig:setup} b). 

Detecting the signal cone ($\mvec{k_{s}}$) results in circular scattering patterns that are characteristic of XPDC and provide access to the underlying high-energy polariton.
To image these signatures, we employ a momentum resolved detection scheme analogous to our previous work \cite{krebs_x-ray_2025}.
As visualized in Fig.~\ref{fig:setup} c, the scheme centers on a spherically-bent crystal analyzer (SBCA) used in combination with a 2D pixel detector.
The signal cone is Bragg-reflected by the SBCA, which refocuses it and simultaneously acts as an energy filter. Here, we use a Si(660) SBCA, bent at 1 m radius and operating at a fixed backscattering energy of $\hbar \omega_s = 9.69~\text{keV}$. To image the energy-filtered signal, we position the detector before the analyzer’s focal plane and are, thus, able to record the full angular distribution of scattered photons \cite{krebs_x-ray_2025}.
In this scheme, we denote the in-plane scattering angle among $\mvec{k_{p}}$ and $\mvec{k_{s}}$ by $2\theta$ (shown vertically on the detector image) and the out-of-plane angle by $\chi$ (shown horizontally).
While the energy $\hbar \omega_s$ is fixed by the SBCA, we can adjust the pump photon energy $\hbar \omega_p$ via the incident monochromator (here, a Si(111) double crystal monochromator). Effectively, this allows us to tune the idler energy as $\omega_i = \omega_p - \omega_s$.
Thereby, we are able to explore the spectral properties of the polariton that is launched through XPDC. In addition to spectral studies, angular scans of the phase-matching condition (cf. Fig.~\ref{fig:setup} b) are also possible.
To this end, the sample is rotated in the rocking angle $\Omega$, which tilts the associated reciprocal lattice vector $\mvec{G}$. The described setup was implemented at the beamline ID20 of the ESRF, making use of its x-ray Raman spectrometer \cite{huotari_large-solid-angle_2017} to perform the reported experiments.
We want to remark that the process of polaritonic XPDC can, in fact, also be thought of as non-resonant x-ray Raman scattering on polaritonic excitation modes - in line with optical Raman scattering (cf. Refs. \cite{henry_raman_1965,chekhova_study_1993}).
In this picture, the Raman process transfers energy $\Delta E = \hbar(\omega_p - \omega_s)$ and momentum $\mvec{Q} = \mvec{k_p} - \mvec{k_s}$ onto the sample, which can excite a polariton, if its phase-matching condition is met.
In contrast to the typical dispersion-relations of other excitations in x-ray Raman scattering (e.g., plasmons \cite{sahle_planning_2015}), the phase-matching condition is very sharply defined.

\subsection*{Measurements of polaritonic scattering}
\begin{figure}[h!]
\centering
\includegraphics[width=0.8\linewidth]{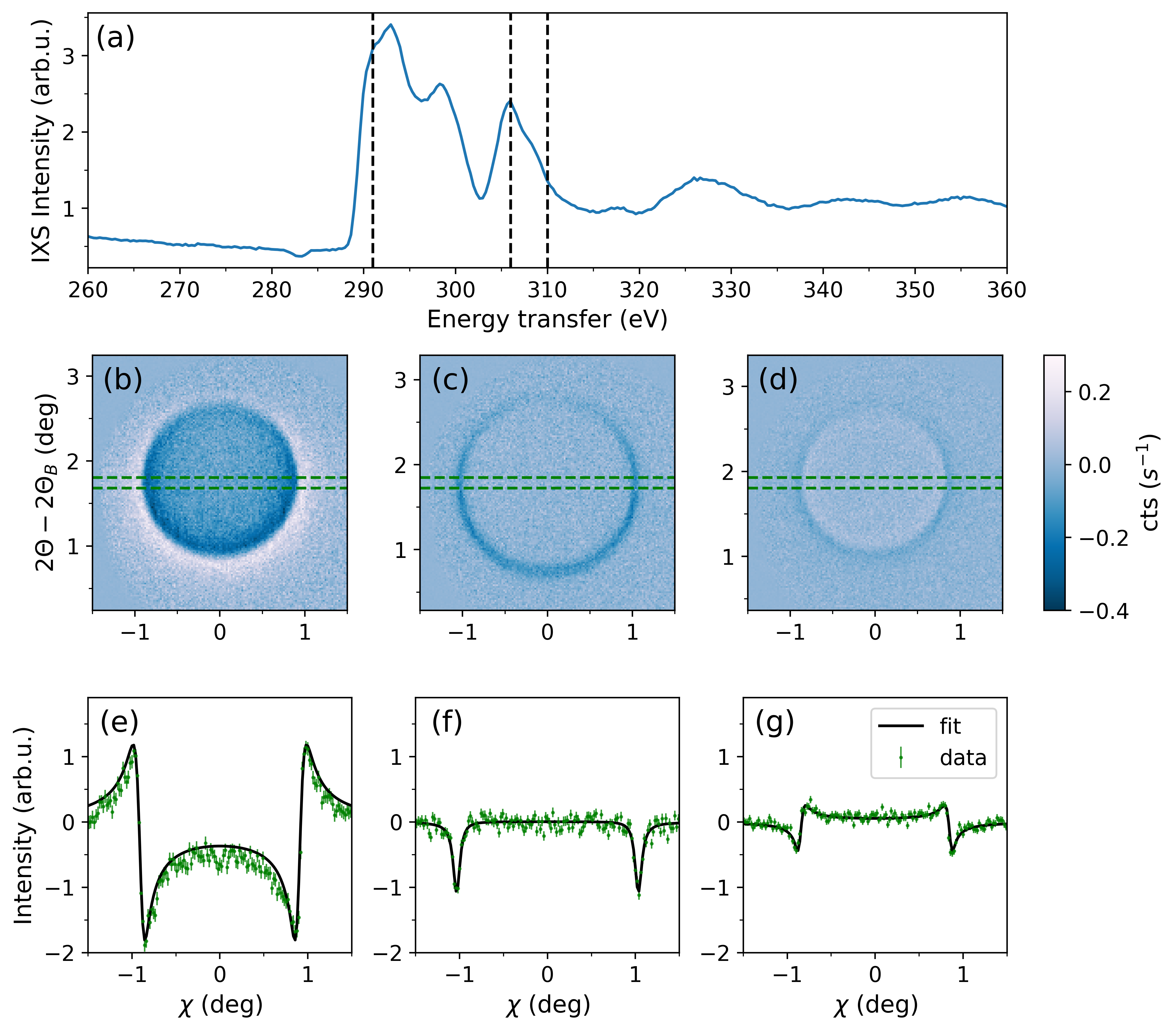}
\caption{Spectral features of polaritonic scattering: (a) Inelastic x-ray scattering signal measured around the Carbon K-shell ionization edge. Characteristic XPDC scattering signatures (b) directly at the carbon K-edge at 291~eV, (c) at the third spectral peak at 306~eV and (d) behind all pronounced edges at 310~eV. Selected energies are indicated by vertical dashed lines in (a). The patterns exhibit pronounced variation in their radial intensity modulation as well as slight variations in size, depending on their phase-matching geometry, idler energy and refractive index. Summation over the line-outs delineated by green dashed borders in (b - d) yield intensity profiles (e - g), which exhibit strongly varying features: (e) two pronounced outer peaks, opposed by inner dips and a lowered plateau; (f) two sharp negative features; (g) two outer dips, opposed by inner peaks and an elevated plateau. All features can be fitted using the TLS model described in Eq.~\ref{eq:ratef} (black lines, in e - g).}
\label{fig:IXS-vs-XPDC}
\end{figure}

In order to explore the spectral behavior of the newly-accessible high-energy polaritons\cite{krebs_x-ray_2025}, we choose diamond as the (nonlinear) sample material.
In addition to its high crystalline quality and outstanding hardness to radiation, diamond offers a richly structured spectrum of excitations around its K-shell ionization edge, which are amenable to polaritonic hybridization. Prior to dedicated XPDC detection, we assess this spectral region from 260~eV to 360~eV with an inelastic x-ray scattering (IXS) measurement (Fig.~\ref{fig:IXS-vs-XPDC} a). Capturing the diffuse IXS signal filtered by the SBCA at $\hbar \omega_s = 9.69~\text{keV}$, we vary the incident photon energy ($\hbar \omega_p$) to scan the effective energy transfer. The resultant IXS spectrum shows the three characteristic K-shell ionization (energy loss) peaks of diamond at 294 eV, 299 eV and 306 eV alongside well-known post-edge structures \cite{galambosi_symmetry_2007,morar_observation_1985}. The energy resolution obtained from the overall setup during this scan is 1.25~eV (FWHM).

Subsequently, we proceed to measure XPDC at several positions throughout this spectrum - adjusting the rocking angle ($\Omega$) of the diamond for phase-matching and using the momentum-resolving capabilities of our setup to capture the characteristic XPDC cone. Figures~\ref{fig:IXS-vs-XPDC} b-d show three examples of the resulting circular scattering signatures: (b) directly at the carbon K-edge at 291~eV, (c) at the third spectral peak at 306~eV and (d) behind all pronounced edges at 310~eV (all positions are indicated by dashed lines in Fig.~\ref{fig:IXS-vs-XPDC} a). The adjusted rocking angles for these cases are $\Delta\Omega = 3.60~\text{deg}, 3.70~\text{deg and } 3.87~\text{deg}$, respectively, with each being measured relative to the Bragg condition for the pertaining pump energy (i.e., $\Omega = \theta_B^\text{pump} + \Delta\Omega$). A full list of all measured cases is given in Methods Sec.~\ref{sec:meth_data}.
The images shown in Fig.~\ref{fig:IXS-vs-XPDC} b-d are generated from signal data by subtracting a background acquisition, which is taken close to - yet outside - the phase-matching ellipse (for each case: $\Omega + 0.7~\text{deg}$). Thus it contains no XPDC signal, but compensates for other background scattering effects and instrumental artifacts. The observed circular signatures vary slightly in size, depending on their phase-matching geometry, idler energy and refractive index (see below). Moreover, they vary markedly in their patterns of (radial) intensity distribution, which becomes even more apparent when taking lineouts across the center of the circles. Summing over the 7 pixels framed by green dotted lines in Figs.~\ref{fig:IXS-vs-XPDC} b-d (i.e., ranges of $\Delta2\theta = 0.143 ~\text{deg}$ each), we obtain the intensity profiles of Fig.~\ref{fig:IXS-vs-XPDC} e-g. Directly at the K-edge at 291~eV, the lineout (Fig.~\ref{fig:IXS-vs-XPDC}~e) shows two pronounced outer peaks that are opposed by inner dips and a lowered plateau around the origin $\chi=0$. At 306~eV idler energy, the lineout (Fig.~\ref{fig:IXS-vs-XPDC}~f) solely exhibits two sharp negative features, whereas the measurement at 310~eV (Fig.~\ref{fig:IXS-vs-XPDC}~g) again resumes Fano-like shapes \cite{fano_effects_1961,krebs_x-ray_2025} similar to Fig.~\ref{fig:IXS-vs-XPDC}~e, but sharper and with flipped signs. Furthermore, the last case also shows a central plateau, which is slightly elevated this time rather than lowered as in Fig.~\ref{fig:IXS-vs-XPDC}~e. These rich variations are imprints of the underlying high-energy polariton and can be analyzed and understood in terms of the theoretical model which we present in the following section.

\begin{figure}[h!]
\centering
\includegraphics[width=0.8\linewidth]{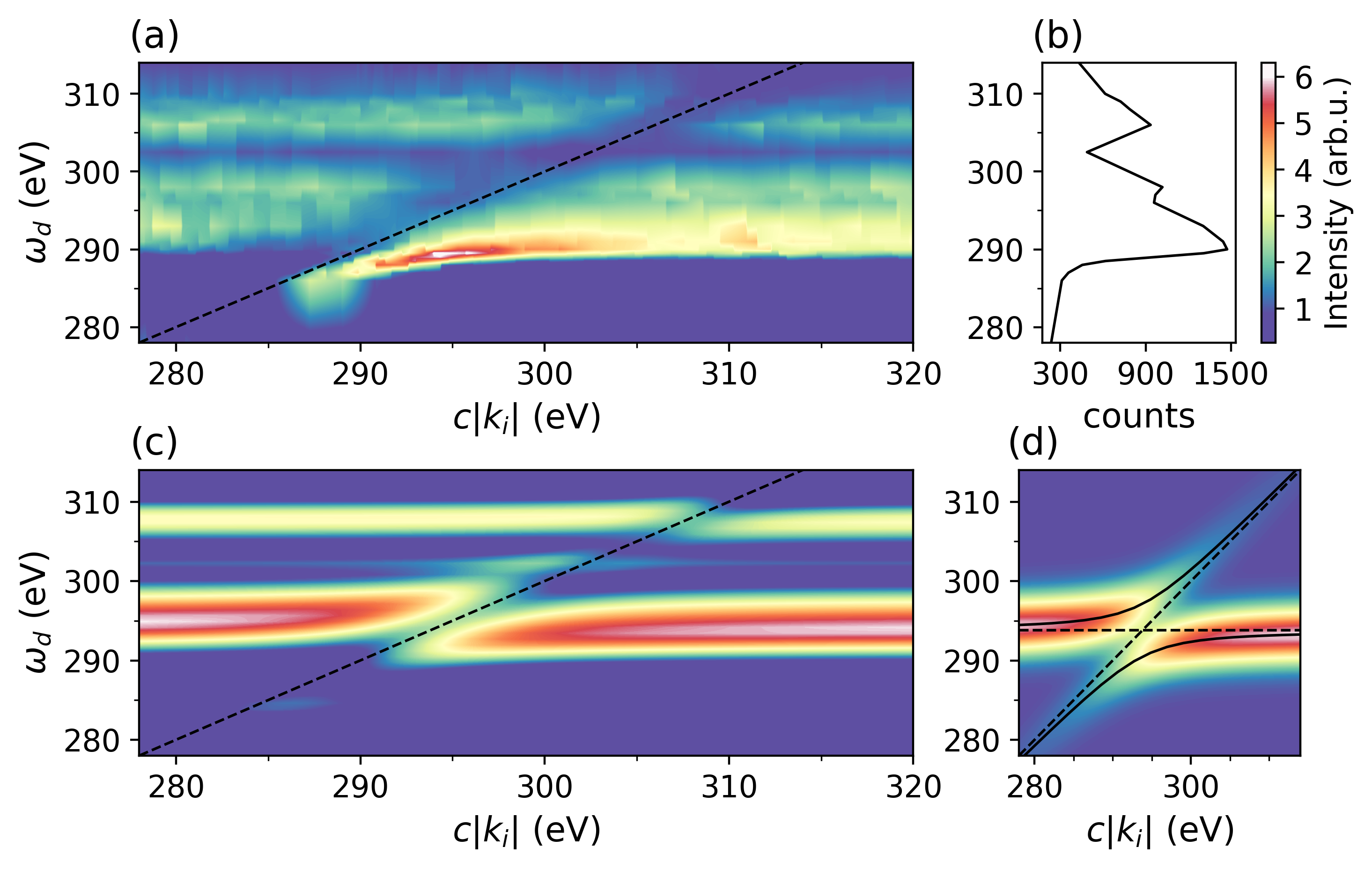}
\caption{2D polariton spectral maps: (a) Central lineouts of the experimental scattering signal (without subtracting the background) are plotted against the energy transfer (or detection energy) $\omega_d = \omega_p - \omega_s$ in one dimension (vertical axis) and the effective idler momentum $c | \mvec{k_i} |$ rescaled to energy in the other dimension (horizontal axis). The map shows a distinct nodal line across the diagonal (black dashed line is overlayed as guide to the eye), which is characteristic of the polariton's anti-crossing dispersion branches. (b) Integration along the horizontal axis effectively yields the IXS signal (cf. Fig.~\ref{fig:IXS-vs-XPDC} a). (c) Simulation of the polariton spectral map based on our two-level system (TLS) model for a range of excitation energies, fit to experimental data (cf. Methods Sec.~\ref{sec:meth_data}). The simulated map shows excellent agreement with measurements (a), replicating the diagonal node as well as the so-called second band-gap ($\omega_d \approx 302.5 ~\text{eV}$). (d) Illustration for a single TLS, showing the direct correspondence of the spectral map to the polaritonic dispersion scheme of the TLS (black overlay). Uncoupled photonic (diagonal dashed) and electronic level (horizontal dashed) hybridize into lower and upper polariton branch (curved, black lines). }
\label{fig:polariton-maps}
\end{figure}

In order to expose the polaritonic nature of this underlying excitation particularly clearly, we introduce a new representation of the measured XPDC data in the form of a 2D spectral map (see Fig.~\ref{fig:polariton-maps}~a). This allows us to visualize the polaritonic dispersion branches directly and observe their characteristic anticrossing. Our procedure is inspired by 2D spectroscopy techniques, prominently RIXS \cite{weinhardt_resonant_2009} or Raman-scattering \cite{doi:10.1366/0003702934067694}, and naturally extends ideas of light-scattering by polaritons \cite{huang_lattice_1951,henry_raman_1965,aktsipetrov_frequency-angle_nodate,kulevsky_light_1975} or polariton k-spectroscopy \cite{chekhova_study_1993,coffinet_coherent_1969} from the optical domain.
To create a polariton spectral map in 2D, we plot cuts of the scattering signal (without subtracting background) against the analyzed energy transfer or detection energy $\omega_d = \omega_p - \omega_s$ in one dimension (vertical axis) and the effective idler photon momentum (scaled to energy) $c | \mvec{k_i} |$ - derived from the locally phase-matched wavevector - in the other dimension (horizontal axis). The polariton map of Fig.~\ref{fig:polariton-maps} a is created by combining scans with detection energies between 278~eV and 314~eV (see Methods Sec.~\ref{sec:meth_data} for full list), effectively stacking the lineouts that were introduced by transitioning from Figs.~\ref{fig:IXS-vs-XPDC}~b-d to e-g on their positive half-side ($\chi \ge 0$). The horizontal coordinate is mapped from out-of-plane scattering angle $\chi$ to momentum (for details, see Methods Sec.~\ref{meth-conv}), with a linear interpolation being applied to achieve an equidistant grid for plotting. 

In the resulting 2D map, we observe the scattering signal to be almost absent for detection energies $\omega_d$ below the carbon K-edge (289.5~eV) except for a weak tail around $c | \mvec{k_i} | \approx 287~\text{eV}$. Above the edge, a modulated scattering signal fills most of the spectral map - with two prominent exceptions: 1. The 2D map is crossed by a distinct nodal line along the diagonal, which marks the essential anti-crossing of all polaritonic dispersion branches in the system. 2. There is a significant suppression of the scattering signal all along the horizontal line at $\omega_d = 302.5 ~\text{eV}$. This corresponds to the pronounced dip in the IXS spectrum at the same energy (cf. Fig.~\ref{fig:IXS-vs-XPDC} a). In fact, if we sum the 2D spectral map along the horizontal axis ($c | \mvec{k_i} |$), we retrieve said IXS spectrum, which is demonstrated by the marginal Fig.~\ref{fig:polariton-maps} b. 
Retaining the second dimension, however, we can illustrate the direct correspondence of the polariton map to the dispersion scheme of a simple polaritonic two-level system (TLS). This is illustrated in its simplest form in Fig.~\ref{fig:polariton-maps} d for a hybridizing excitation at $\omega_e = 293.80 ~\text{eV}$, mixing with the photonic dispersion (diagonal, dashed black line) to give two polaritonic dispersion branches. Building on this correspondence, we present more detailed theoretical modelling in the next section, which allows us to represent the full scattering signal using multiple independent TLS (Fig.~\ref{fig:polariton-maps} c).
Despite the model’s simplicity, the simulated polariton map already shows remarkable similarity to the experimental results - reproducing the clearly visible anticrossing at the expected diagonal (dashed black line as guide to the eye), the spectral depression at $\omega_d = 302.5 ~\text{eV}$ and the gentle curvature of the dispersion branches. Besides this, one notable deviation remains, which concerns the experimental asymmetry of the scattered intensity near the K-edge from left to right, i.e., between the upper and lower dispersion branch pertaining to the hybridizing level at $\omega_e = 293.80 ~\text{eV}$. Its more accurate description goes beyond the scope of our TLS modelling and is not the subject of this initial exploration.

\subsection*{Theoretical modelling of polaritonic XPDC}
For our present discussion, we make use of simple, yet effective two-level system (TLS) models to help understand the spectral behavior of high-energy polaritons.
As the maximally reduced model for hybridization, the TLS has found wide-spread application to polaritonic phenomena \cite{mandal_theoretical_2023,torma_strong_2015,ebbesen_hybrid_2016,toffoletti_coherent_2025,baranov_novel_2018}, where generically one of its levels is taken to represent a photonic excitation (e.g., cavity photon) and the other level is used to model the excitation of its material counterpart (e.g., phonon, surface-plasmon or exciton \cite{basov_polariton_2020}).
Variations of this can see two coupled TLS - one for light and matter each - or multi-mode assemblies of several TLS and harmonic oscillator modes \cite{mandal_theoretical_2023,baranov_novel_2018,blaha_beyond_2022}. In our case, we focus on the singly excited subspace of the polariton and identify one basis state of the TLS with the photonic, idler component $\left|\phi_\gamma\right\rangle={(1,0)}^T$ and the other with the electronic component $\left|\phi_e\right\rangle={(0,1)}^T$ of the polariton. The latter, specifically, corresponds to an electronic excitation from the carbon K-shell, for which we will subsequently trace the spectral behavior across its ionization edge (cf. Fig.~\ref{fig:IXS-vs-XPDC} a). The two states hybridize with an effective coupling strength V introduced via the following Hamiltonian 
\begin{equation}\label{ham}
    H^{\text{pol}} = \hbar \begin{pmatrix}
\omega_\gamma & 0\\
0 & \omega_e
\end{pmatrix}
+
\begin{pmatrix}
0 & V\\
V^* & 0
\end{pmatrix}
\end{equation}
where the diagonal entries mark the bare states’ energies. After diagonalizing (Eq.~\ref{ham}), we obtain the well-known polaritonic eigen-energies
\begin{equation}\label{enpol}
    E_{\pm} = \Bigg( \frac{\hbar (\omega_\gamma + \omega_e)}{2} \pm \sqrt{\frac{\hbar^2}{4}(\omega_\gamma - \omega_e)^2 + |V|^2}\Bigg).
\end{equation}
As a function of the photonic energy $\omega_\gamma$, these two branches $E_{\pm}$ exhibit the characteristic anti-crossing around the nodal point $\omega_e = \omega_\gamma$. This is illustrated in Fig.~\ref{fig:polariton-maps} d - for parameters $\omega_e = 293.80$~eV and $V = 3.32$~eV matching the hybridizing excitation with the highest coupling strength of the spectral map Fig.~\ref{fig:polariton-maps} c. 

Going one step further, we can simulate the full polaritonic scattering signal, if we embed this TLS model in a suitably adapted IXS description\cite{krebs_x-ray_2025}. 
To this end, we start from the double-differential scattering cross section of inelastic x-ray scattering\cite{schulke_electron_2007}
\begin{equation}
\frac{\text{d} \sigma}{\text{d}\Omega_s \text{d}\omega_s}  =  \Big( \frac{\text{d} \sigma}{\text{d}\Omega_s} \Big)_\text{Th} ~
  \frac{\omega_s}{\omega_p} ~~
  S(\mvec{Q},\omega)  
  \label{eq:IXS-gen}
\end{equation}
with $\Big( \frac{\text{d} \sigma}{\text{d}\Omega_s} \Big)_\text{Th}$ denoting the Thomson scattering cross section and $S(\mvec{Q},\omega)$ the dynamic structure factor for inelastic scattering. 
Note that for the remainder of this section, we adopt the system of atomic units for more concise notation, thereby omitting factors $\hbar$, for instance.
Here, we introduce the polaritonic model by prescribing that the time-evolution within the electronic correlator should (only) proceed via the two effective states of the TLS:
\begin{align}
S(\mvec{Q},\omega) =& \int \diff \mvec{x} \int \diff \mvec{x'} e^{-i \mvec{Q}(\mvec{x}-\mvec{x'})} \int \diff t e^{\iu \omega t} \langle n(\mvec{x},t) n(\mvec{x'},0) \rangle \label{l1}\\
\approx& S^{\text{pol}}(\mvec{Q},\omega) \equiv \int \diff \mvec{x} \int \diff \mvec{x'} e^{-\iu \mvec{Q}(\mvec{x}-\mvec{x'})} \int \diff t e^{\iu \omega t} \langle \phi_I | n(\mvec{x},0) \sum_\phi^{\text{2lvl}} |\phi\rangle \langle \phi| U(t,0) \sum_{\phi'}^{\text{2lvl}} |\phi'\rangle \langle \phi'| n(\mvec{x}',0) |\phi_I\rangle 
\end{align} 
This expression can be further simplified to the sole contribution of $|\phi_e \rangle$, since only this electronic component of the polariton’s bare states can couple via the electron density operator $n(\mvec{x})$ and contribute to $\langle \phi| n |\phi_I \rangle$. The other component ($|\phi_\gamma \rangle$) would instead require the simultaneous creation of a photon. In the resulting expression
\begin{align}
S^{\text{pol}}(\mvec{Q},\omega) = \int \diff \mvec{x} \int \diff \mvec{x'} e^{-\iu \mvec{Q}(\mvec{x}-\mvec{x'})} \langle \phi_I | n(\mvec{x},0) | \phi_e \rangle \langle \phi_e | n(\mvec{x},0) | \phi_I \rangle \int \diff t e^{\iu \omega t} \langle \phi_e | U(t,0) \phi_e \rangle 
\end{align}
we make use of the diagonal representation of $U$, or $H^{\text{pol}}$ (Eq.~\ref{ham}), respectively, to solve the time-evolution via the eigenenergies (Eq.~\ref{enpol}) and corresponding eigenstates
\begin{equation}
    |\mathbf{\phi}_\pm^{\text{pol}} \rangle = c_\pm
    \left( \begin{array}{c}
    1\\
    \frac{-V^*}{ \omega_\gamma - E_\pm}\\
    \end{array} \right)
\ \ \ \ \ \text{with} \ \ \ \ \ 
    c_\pm = \frac{1}{\sqrt{1 + |V|^2/( \omega_\gamma - E_\pm)^2}}.
\end{equation}
Accounting for the basis transformation $| \phi_e \rangle \rightarrow |\phi_\pm^{pol} \rangle$ in terms of $c_\pm$, and evaluating the temporal Fourier transform for an energy conserving delta-function, we arrive at the polaritonically modified dynamic structure factor
\begin{align}\label{spol2}
S^{pol}(\mvec{Q},\omega) =& \underbrace{|\int \diff \mvec{x} e^{\iu \mvec{Q} \mvec{x}} \langle \phi_I | n(\mvec{x}) | \phi_e  \rangle|^2}_{\equiv S(\mvec{Q})} \sum_\pm |c_\pm|^2 \delta(E_\pm - \omega).
\end{align}
Here, we include all remaining scattering effects within a static prefactor $S(\mvec{Q})$, implying no more spectral dependence across the narrow energy window of our detection.

\begin{figure}[ht]
\centering
\includegraphics[width=0.9\linewidth]{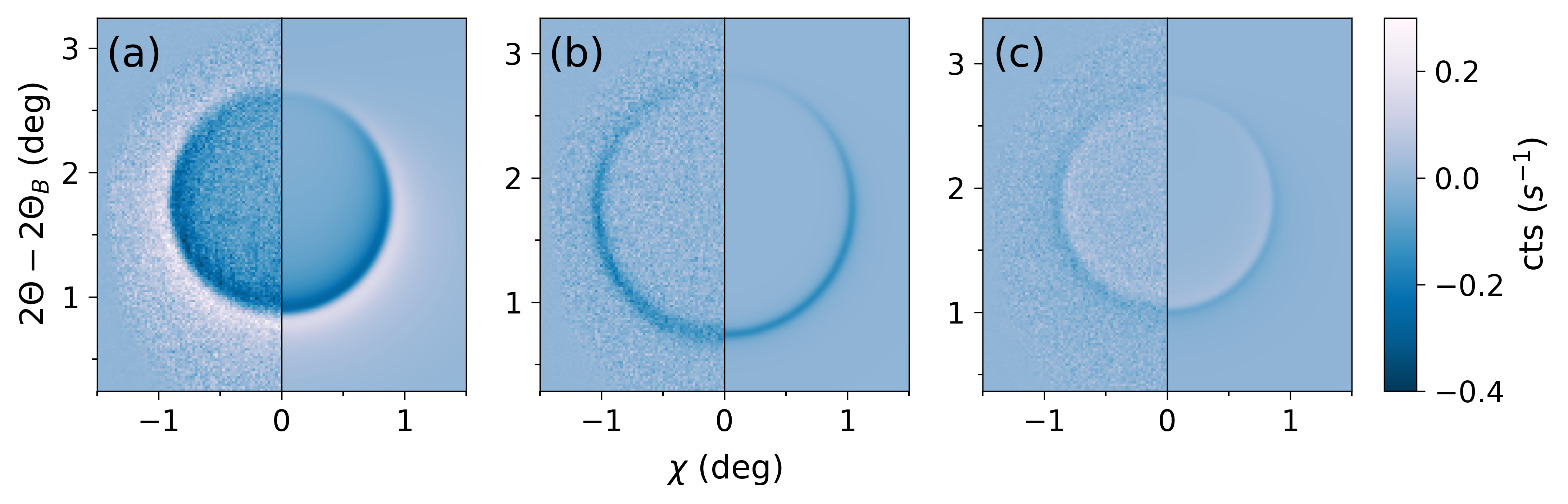}
\caption{Comparison of experimental (left) and simulated (right) scattering signatures for different polaritonic energies around the Carbon K-edge at (a) $291$~eV, (b) $306$~eV and (c) $310$~eV (cf. Fig.~\ref{fig:IXS-vs-XPDC}). The experimental and simulated data are in excellent agreement for cone-diameters, sign of the signal and intensity modulation. The simulation is based on the effective TLS model as described in this section. } 
\label{fig:simulation}
\end{figure}

In order to compare these results directly to the real scattering signal (e.g., Figs.~\ref{fig:IXS-vs-XPDC} b-d), we convert the cross section (Eq.~\ref{eq:IXS-gen}) into a count rate $R$ for our setup, integrating it with a spectrally resolved model of the pump photon flux $J_p(\omega_p)$ and a spectral transmission function $T_s(\omega_s)$ that models our analyzer:
\begin{align}
  R
  &\equiv
  \int_\text{pix} \, \diff \Omega_s \int \! \diff \omega_s \int \! \diff \omega_p ~
  T_\text{s}(\omega_s) ~ J_\text{p}(\omega_p) ~
  \frac{\diff \sigma}{\diff \Omega_s \diff \omega_s}  \\
   &\approx \int_\text{pix} \, \diff \Omega_s \int \! \diff \omega_s ~ \int \! \diff \omega_p ~
  T_\text{s}(\omega_s) ~ J_\text{in}(\omega_p) ~
  \Big( \frac{\diff \sigma}{\diff \Omega_s} \Big)_\text{Th} ~ 
  \frac{\omega_s}{\omega_p} 
  \cdot S(\mvec{Q}) \sum_\pm |c_\pm|^2 \delta(E_\pm - (\omega_p - \omega_s))
\end{align}
We further adopt the commonly-used approximation that $\omega_s/\omega_p \approx 1$ and assume that the angular variation of the signal is small around the mean of each pixel on our detector, i.e. $S(\mvec{Q}) \approx S(\mvec{\bar{Q}})$. This reduces the angular integral to a simple function of the Thomson cross section, which can be evaluated numerically as
\begin{align}
  R
  &\approx \Big( \frac{\diff \sigma}{d\Omega_s} \Big)_\text{Th} \Big|_{\Omega_\text{pix}}
   S(\mvec{\bar{Q}}) 
   \int \! \diff \omega_s ~ \int \! \diff \omega_p ~
  T_\text{s}(\omega_s) ~ J_\text{p}(\omega_p) ~
  \sum_\pm |c_\pm|^2 \delta(E_\pm - (\omega_p-\omega_s)).
\end{align}
Finally, we render the description of the pump beam $J_p(\omega_p)$ and the transmission function of the analyzer $T_s(\omega_s)$ explicit by modelling them as Gaussians with bandwidths $\Gamma_p$ and $\Gamma_d$ around center energies $\bar{\omega}_p$ and $\omega_{d}$, respectively: 
\begin{align}
 R
  &= \Big( \frac{\diff \sigma}{\diff \Omega_s} \Big)_\text{Th}|_{\Omega_\text{pix}}
   S(\mvec{\bar{Q}})  \sum_\pm |c_\pm|^2
   \int \! \diff \omega_p ~ 
  J e^{-\frac{-(\omega_p - \bar{\omega}_p)^2}{2 \Gamma_p^2}} ~ T e^{-\frac{-(\omega_p +E_\pm+\omega_d - \bar{\omega}_p)^2}{2 \Gamma_d^2}}  \\
  &= \underbrace{\Big( \frac{\diff \sigma}{\diff \Omega_s} \Big)_\text{Th}|_{\Omega_\text{pix}}
   S(\mvec{\bar{Q}}) J T \sqrt{\frac{\pi}{1/(2\Omega_p^2) + 1/(2\Omega_d^2)}}}_{\equiv p} \sum_\pm |c_\pm|^2
  e^{\frac{-(E_\pm - \omega_d)^2}{2\Gamma_d^2 + 2\Gamma_p^2}}
\end{align}
The resulting prefactor $p$ is effectively determined by the Compton background and can be extracted from the average background count-rate in practice. The combined spectral broadening is $\Gamma_{\text{setup}}^2 = \Gamma_d^2 + \Gamma_p^2$. On phenomenological grounds, we adapt the TLS to feature decay (cf. Ref.\cite{torma_strong_2015}), i.e., we add an intrinsic bandwidth of the polariton $\Gamma_{pol}$. Thus, we end up with the total width $\Gamma^2 = \Gamma_{\text{pol}}^2 + \Gamma_{\text{setup}}^2$ and the overall signal rate described by
\begin{align}\label{eq:ratef} 
  R= p \sum_\pm |c_\pm|^2
  e^{\frac{-(E_\pm - \omega_d)^2}{2\Gamma^2}}.
\end{align}
Showcasing the versatility of this simple TLS model, we can now use it to reproduce all different scattering signatures found in Figs.~\ref{fig:IXS-vs-XPDC} b-d by corresponding simulations in Figs.~\ref{fig:simulation} a-c. These are obtained using the parameters $\omega_e = 293.80$~eV, $305.84$~eV and $308.27$~eV as well as $\Gamma = 2.40$~eV, $1.90$~eV and $1.59$~eV with the coupling strengths taken to be $V = 3.34$~eV, $2.28$~eV and $1.55$~eV, respectively. In addition, the coupling term is angularly modulated with the polarisation factor
\begin{equation}
    V \xrightarrow[]{} V \cdot \big( 1 - |\hat{k}_\gamma^\text{eff} \cdot \hat{G}|^2 \big)
\end{equation}
which derives from the transversality constraint on the XPDC idler photon \cite{boemer_towards_2021,krebs_x-ray_2025}.
Finally, the prefactor p is fixed to match the experimental background away from the phase-matching condition. In terms of Eq.~\ref{eq:ratef}, this corresponds to taking the limit $\omega_\gamma \xrightarrow[]{} \infty$, such that the photonic excitations cannot fulfill momentum conservation and the scattering expression reduces to a single term $R \xrightarrow[]{} R_{\text{average}} = p \cdot \exp{(\frac{-(\omega_d - \omega_c)^2}{2 \Gamma^2})}$. Here, it takes on the values $p = 0.43,~0.31$ and $0.17$ counts/(pixel*second), respectively.

\subsection*{Spectral evaluation of experimental results}
Beyond simulating individual XPDC patterns, we can apply our theoretical model to systematically fit all experimental data (cf. listing in Methods Sec.~\ref{sec:meth_data}). To this end, we employ Eq.~\ref{eq:ratef} as the forward model for optimization, which allows us to extract essential parameters of the underlying TLS and study their spectral dependence. The resulting insights contribute further to our understanding of the high-energy polariton and suggest future applicability for material diagnostics.

\begin{figure}[ht]
    \centering
    \includegraphics[width=\linewidth]{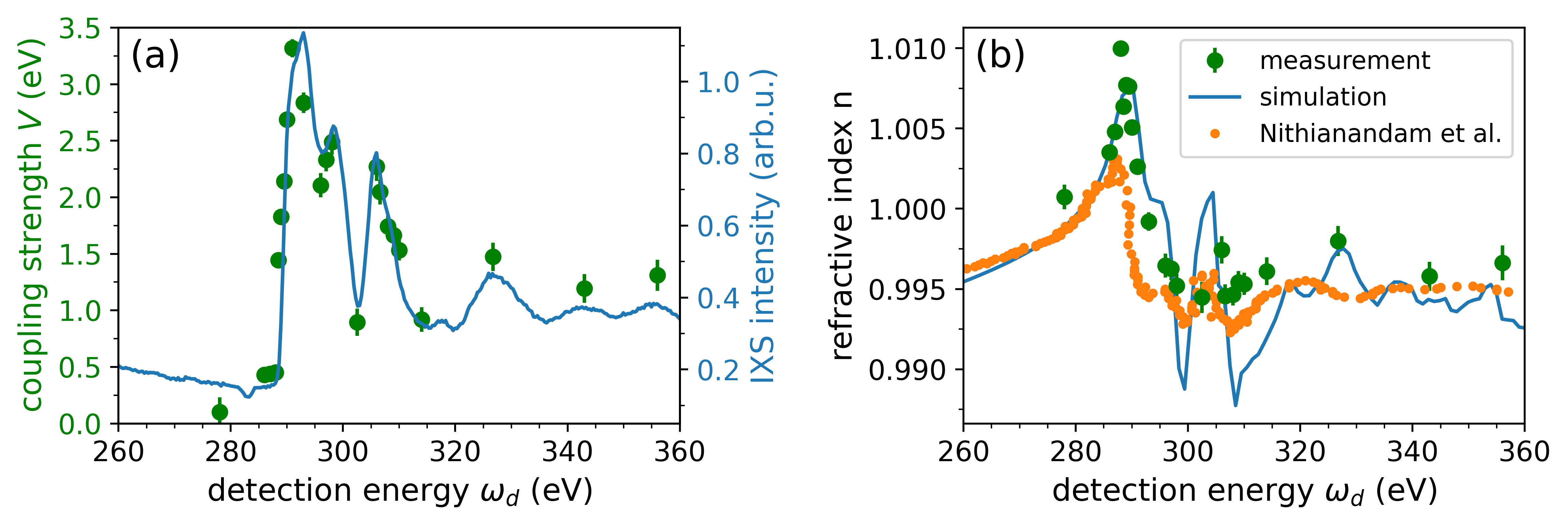}
    \caption{Spectral dependence of polaritonic parameters: 
a) The coupling strength $V$ extracted via the TLS model (green points with  error-bars) for different detection energies $\omega_d = \omega_p - \omega_s$ closely follows the scaled IXS signal (blue line).
b) The measured refractive index $n$ at the K-edge in diamond (green points with error bars) for varying detection energies $\omega_d$ agrees well with simulated data from DFT calculations (blue line). Both curves exhibit systematically higher values of the refractive index than previous data obtained by Nithianandam et al. using Kramers-Kronig inversion (orange dots, adapted from Ref.~\cite{nithianandam_synchrotron_1993}).
Experimental error estimates are based on counting error (standard deviation of background fluctuations) and propagated via the covariance matrix of the model fit (square root of diagonal elements).}
        \label{fig:spectral}
\end{figure}

Considering the polaritonic coupling strength $V$ first, we find it to closely follow the IXS signal measured before (see Fig.~\ref{fig:spectral} b). At the peak of the absorption profile (i.e., 291~eV), we observe the highest coupling strength, whereas before (< 289.5~eV) and behind (> 310~eV) the edge, $V$ drops significantly alongside the IXS signal. In fact, both spectra trace the density of states for dipole-allowed transitions (here: p-DOS \cite{galambosi_symmetry_2007,PhysRevB.4.3610}) - with the polaritonic signal being sensitive even to the dip at 302.5~eV, which is associated with the second band gap of diamond \cite{galambosi_symmetry_2007,PhysRevB.4.3610}. The close correspondence of polaritonic XPDC and IXS signal strength also mirrors earlier observations by Tamasaku et al., who analyzed their respective XPDC measurements using a phenomenologically motivated Fano formula \cite{tamasaku_determining_2009}. 
Our new, polaritonic interpretation of XPDC further allows us to assess the underlying light-matter interaction with regard to the strong-coupling criterion $2V \ge \hbar\Gamma$ \cite{ebbesen_hybrid_2016,garcia-vidal_manipulating_2021,sanchez-barquilla_theoretical_2022,mandal_theoretical_2023,li_molecular_2022,torma_strong_2015}. The respective overall bandwidth $\Gamma$ follows a similar spectral trend as $V$, likewise peaking at $\omega_d = 291 ~\text{eV}$ for a value of $\hbar\Gamma = 2.44 ~\text{eV}$. The resulting ratio $2V/\hbar\Gamma = 2.72$ places the observed high-energy polariton at the K-edge firmly within the strong-coupling regime - far exceeding the case studied previously in the EUV-regime \cite{krebs_x-ray_2025}. This higher coupling strength and commensurately improved signal opens opportunities to study strong-coupling phenomena at soft x-ray energies with relative ease in the future.

As a macroscopic consequence of this strong light-matter interaction, spectral structure is also imprinted onto the refractive index of the diamond studied by XPDC (see Fig.~\ref{fig:spectral} b). Using the dispersion relation $\omega_\gamma = c | \mvec{k_i} | / n$ within our TLS model, the refractive index can be extracted alongside other model parameters from fits to the experimental data (cf. listing in Methods Sec.~\ref{sec:meth_data}). The overall modulation of $n$ reaches at most $\approx 1\%$, nonetheless, our analysis can trace its spectral dependence with high sensitivity and very little error (estimate indicated by error bars in Fig.~\ref{fig:spectral} b). Notably, direct experimental data on bulk refractive indices is exceedingly difficult to obtain in this photon energy regime, especially at the presented level of resolution. As the closest experimental reference for diamond, we compare our results to Kramers-Kronig (KK) inverted reflectance data extracted from Ref. \cite{nithianandam_synchrotron_1993} (orange dots in Fig.~\ref{fig:spectral} b). While their KK transform appears to be systematically smaller than our measurements, the observed spectral features are in very good agreement. For further benchmarking of our results, we also compare to theoretical data obtained from DFT simulations. Using the all-electron code FHI-aims \cite{blum_ab_2009,ren_resolution--identity_2012} to access the carbon K-edge in diamond, we extract the refractive index from HSE06 hybrid potential calculations \cite{heyd_hybrid_2003}.
The resulting curve (blue line in Fig.~\ref{fig:spectral} b) matches the features of our experimentally determined refractive index well and supports the higher overall magnitude of $n$ compared to Ref. \cite{nithianandam_synchrotron_1993}. These first findings highlight the potential of polaritonic XPDC to access soft x-ray refractive indices at an unprecedented level of detail, as future refinements of setup and measurement bandwidth will be implemented.  

\section*{Discussion}

Following the recent observation of EUV-polaritons through x-ray parametric down-conversion \cite{krebs_x-ray_2025}, we have investigated the spectral properties of analogous high-energy polaritons at the carbon K-edge in diamond. In the process, we introduce a 2D spectral map that visualizes the polaritonic dispersion directly from the nonlinear scattering signal and match it with an effective theoretical model for further analysis. We find that the hybridization at the absorption edge results in a substantially higher coupling strength than the previously reported EUV scenario, reaching well into the strong-coupling regime. We expect this effect to be even more pronounced on isolated pre-edge resonances as can be found in graphite or hexagonal Boron Nitride, for instance \cite{schulke_interband_1988,mcdougall_near_2014}. Here, the access to clean few-level systems, would allow for the unprecedented production and manipulation of non-classical hybrid states at soft x-ray wavelengths. Moreover, we demonstrate that the spectral sensitivity of polaritonic XPDC can be utilized to measure refractive properties of the bulk sample at high resolution. This presents a rare opportunity in the EUV to soft x-ray regime, which is otherwise restricted to (near) surface probes, such as reflectometry \cite{soufli_reflectance_1997,dorney_actinic_2024}. Knowledge on optical constants for EUV-lithography and fundamental material science \cite{tallents_lithography_2010,wagner_lithography_2010,ciesielski_determination_2022,wood_improved_2014,sokolov_at-wavelength_2016,behringer_characterization_2015} can thus prospectively be complemented by measuring the polaritonic down-conversion of x-rays.

\section*{Methods}


\subsection{Fitting Parameter table} 
\label{sec:meth_data}

Below, we tabulate all relevant polaritonic parameters extracted from fitting our TLS model to the measured XPDC signals.
\begin{table}[h!]
\centering
\begin{tabular}{cccccc}
\toprule
Detection Energy & Excitation energy & Coupling strength & Bandwidth & Refractive index & Goodness of fit\\ $\omega_{d}$~(eV) & $\omega_{c}$~(eV) & $V$~(eV) & $\Gamma$~(eV) & $n$ & $\chi^2$ \\
\midrule
278 & 284.24 & 0.10 & 1.49 & 1.0007 & 1.36 \\
286 & 290.76 & 0.43 & 1.31 & 1.0035 & 1.31 \\
287 & 291.84 & 0.44 & 1.30 & 1.0048 & 1.01 \\
288 & 293.68 & 0.45 & 1.48 & 1.0100 & 2.00 \\
288.5 & 294.15 & 1.44 & 1.82 & 1.0064 & 1.97 \\
289 & 293.49 & 1.83 & 1.74 & 1.0077 & 1.50 \\
289.5 & 293.15 & 2.14 & 1.71 & 1.0076 & 1.84 \\
290 & 293.07 & 2.69 & 1.91 & 1.0051 & 1.48 \\
291 & 293.80 & 3.32 & 2.44 & 1.0026 & 1.41 \\
293 & 294.09 & 2.83 & 2.07 & 0.9992 & 1.53 \\
296 & 296.55 & 2.11 & 1.75 & 0.9965 & 1.22 \\
297 & 297.57 & 2.33 & 1.91 & 0.9963 & 1.12 \\
298 & 298.26 & 2.49 & 1.94 & 0.9952 & 1.40 \\
302.5 & 302.26 & 0.90 & 0.92 & 0.9945 & 1.06 \\
306 & 305.84 & 2.27 & 1.92 & 0.9974 & 1.60 \\
306.6 & 305.91 & 2.05 & 1.78 & 0.9946 & 1.32 \\
308 & 307.12 & 1.74 & 1.51 & 0.9947 & 1.39 \\
309 & 307.83 & 1.67 & 1.48 & 0.9954 & 1.63 \\
310 & 308.27 & 1.53 & 1.60 & 0.9953 & 1.20 \\
314 & 313.40 & 0.92 & 1.01 & 0.9961 & 1.04 \\
326.7 & 326.84 & 1.47 & 1.48 & 0.9980 & 1.43 \\
343 & 342.86 & 1.19 & 1.26 & 0.9958 & 0.81 \\
356 & 355.51 & 1.31 & 1.60 & 0.9966 & 1.15 \\
\bottomrule
\end{tabular}
\caption{Fit parameters at different idler energies (i.e. detection energies $\omega_{det}$).}
\end{table}

\subsection{Conversion of scattering angle to momenta} 
\label{meth-conv}
Mapping from scattering angles ($2\theta,\chi$) to momentum transfer $\hbar \, \mvec{Q} = \hbar (\mvec{k_p} - \mvec{k_s})$ proceeds in a system of spherical coordinates.
The pump beam defines the polar axis along $\mvec{k_p} =  \omega_p \, c^{-1} \, \mvec{e}_z$, where $c$ is the speed of light in vacuo and $\mvec{e}_z$ the unit vector along z.
The signal photon's $\mvec{k_s}$ is fully determined by its magnitude $| \mvec{k_s}| =  \omega_s \, c^{-1}$ as well as the scattering angles $\theta_s = 2\theta$ for its azimuth and 
\begin{equation}
\chi \rightarrow \phi_s = (\chi - \pi/2) (\sin{2\theta})^{-1} + \pi/2
\end{equation} 
for its polar angle.
Following the phase-matching matching condition illustrated in Fig.~\ref{fig:setup}(b), $\mvec{Q}$ relates to a momentum-, but not necessarily energy-conserving effective idler photon via %
\begin{equation}
\mvec{k_i} = \mvec{Q} + \mvec{G} = \mvec{k_p} - \mvec{k_s} + \mvec{G}.
\end{equation}
Writing the reciprocal lattice vector in the same coordinate system
\begin{align}
\mvec{G} =& \omega_p \, c^{-1} \, \sin{\theta_B} \left(
\begin{array}{c}
0\\
\cos{\theta_B + \Delta\Omega}\\
\sin{\theta_B + \Delta\Omega}
\end{array}
\right),
\label{eq:rec_latt}
\end{align} 
we can compute $\mvec{k_i}$ for all scattering angles. The prefactor in Eq.~\ref{eq:rec_latt} reflects Bragg's law.
We use this mapping with the two-level system to evaluate the pertaining dispersion relations $\omega_i(\mvec{Q})$ or $E_\pm(\omega_i)$ numerically.

\section*{Funding}
This work is supported and partly funded by the Cluster of Excellence “Advanced Imaging of
Matter” of the Deutsche Forschungsgemeinschaft (DFG)—EXC 2056— project ID 390715994 (F.K., A.N., C.B.).

\bibliography{Citationsofspectralpolaritonpaper}

\begin{thebibliography}{10}
\urlstyle{rm}
\expandafter\ifx\csname url\endcsname\relax
  \def\url#1{\texttt{#1}}\fi
\expandafter\ifx\csname urlprefix\endcsname\relax\def\urlprefix{URL }\fi
\expandafter\ifx\csname doiprefix\endcsname\relax\def\doiprefix{DOI: }\fi
\providecommand{\bibinfo}[2]{#2}
\providecommand{\eprint}[2][]{\url{#2}}

\bibitem{bloembergen_nonlinear_1965}
\bibinfo{author}{Bloembergen, N.}
\newblock \emph{\bibinfo{title}{Nonlinear {Optics}: {A} {Lecture} {Note} and
  {Reprint} {Volume}}}.
\newblock Frontiers in physics (\bibinfo{publisher}{W. A. Benjamin},
  \bibinfo{year}{1965}).

\bibitem{franken_generation_1961}
\bibinfo{author}{Franken, P.~A.}, \bibinfo{author}{Hill, A.~E.},
  \bibinfo{author}{Peters, C.~W.} \& \bibinfo{author}{Weinreich, G.}
\newblock \bibinfo{journal}{\bibinfo{title}{Generation of {Optical}
  {Harmonics}}}.
\newblock {\emph{\JournalTitle{Physical Review Letters}}}
  \textbf{\bibinfo{volume}{7}}, \bibinfo{pages}{118--119},
  \doiprefix\url{10.1103/PhysRevLett.7.118} (\bibinfo{year}{1961}).

\bibitem{1967ZhPmR...6..575A}
\bibinfo{author}{{Akhmanov}, S.~A.}, \bibinfo{author}{{Fadeev}, V.~V.},
  \bibinfo{author}{{Khokhlov}, R.~V.} \& \bibinfo{author}{{Chunaev}, O.~N.}
\newblock \bibinfo{journal}{\bibinfo{title}{{Quantum Noise in Parametric Light
  Amplifiers}}}.
\newblock {\emph{\JournalTitle{ZhETF Pisma Redaktsiiu}}}
  \textbf{\bibinfo{volume}{6}}, \bibinfo{pages}{575} (\bibinfo{year}{1967}).

\bibitem{harris_observation_1967}
\bibinfo{author}{Harris, S.~E.}, \bibinfo{author}{Oshman, M.~K.} \&
  \bibinfo{author}{Byer, R.~L.}
\newblock \bibinfo{journal}{\bibinfo{title}{Observation of {Tunable} {Optical}
  {Parametric} {Fluorescence}}}.
\newblock {\emph{\JournalTitle{Physical Review Letters}}}
  \textbf{\bibinfo{volume}{18}}, \bibinfo{pages}{732--734},
  \doiprefix\url{10.1103/PhysRevLett.18.732} (\bibinfo{year}{1967}).

\bibitem{akhmanov_quantum}
\bibinfo{author}{Akhmanov, S.}, \bibinfo{author}{Fadeev, V.},
  \bibinfo{author}{Khokhlov, R.} \& \bibinfo{author}{Chunaev, O.}
\newblock \bibinfo{journal}{\bibinfo{title}{Quantum noise in parametric light
  amplifiers}}.
\newblock {\emph{\JournalTitle{JETP Lett}}} \textbf{\bibinfo{volume}{6}}
  (\bibinfo{year}{1967}).

\bibitem{magde_study_1967}
\bibinfo{author}{Magde, D.} \& \bibinfo{author}{Mahr, H.}
\newblock \bibinfo{journal}{\bibinfo{title}{Study in {Ammonium} {Dihydrogen}
  {Phosphate} of {Spontaneous} {Parametric} {Interaction} {Tunable} from 4400
  to 16 000 Å}}.
\newblock {\emph{\JournalTitle{Physical Review Letters}}}
  \textbf{\bibinfo{volume}{18}}, \bibinfo{pages}{905--907},
  \doiprefix\url{10.1103/PhysRevLett.18.905} (\bibinfo{year}{1967}).

\bibitem{giordmaine_tunable_1965}
\bibinfo{author}{Giordmaine, J.~A.} \& \bibinfo{author}{Miller, R.~C.}
\newblock \bibinfo{journal}{\bibinfo{title}{Tunable {Coherent} {Parametric}
  {Oscillation} in {LiNb} {O} 3 at {Optical} {Frequencies}}}.
\newblock {\emph{\JournalTitle{Physical Review Letters}}}
  \textbf{\bibinfo{volume}{14}}, \bibinfo{pages}{973--976},
  \doiprefix\url{10.1103/PhysRevLett.14.973} (\bibinfo{year}{1965}).

\bibitem{akhmanov1965observation}
\bibinfo{author}{Akhmanov, S.~A.}, \bibinfo{author}{Kovrigin, A.},
  \bibinfo{author}{Piskarskas, A.}, \bibinfo{author}{Fadeev, V.} \&
  \bibinfo{author}{Khokhlov, R.}
\newblock \bibinfo{journal}{\bibinfo{title}{Observation of parametric
  amplification in the optical range}}.
\newblock {\emph{\JournalTitle{JETP Lett}}} \textbf{\bibinfo{volume}{2}},
  \bibinfo{pages}{191} (\bibinfo{year}{1965}).

\bibitem{menzel_photonics_2007}
\bibinfo{author}{Menzel, R.}
\newblock \emph{\bibinfo{title}{Photonics: linear and nonlinear interactions of
  laser light and matter}} (\bibinfo{publisher}{Springer},
  \bibinfo{address}{Berlin}, \bibinfo{year}{2007}), \bibinfo{edition}{2nd ed}
  edn.

\bibitem{li_down-converted_2025}
\bibinfo{author}{Li, B.} \emph{et~al.}
\newblock \bibinfo{journal}{\bibinfo{title}{Down-converted photon pairs in a
  high-{Q} silicon nitride microresonator}}.
\newblock {\emph{\JournalTitle{Nature}}} \textbf{\bibinfo{volume}{639}},
  \bibinfo{pages}{922--927}, \doiprefix\url{10.1038/s41586-025-08662-3}
  (\bibinfo{year}{2025}).

\bibitem{wegner_nif_2004}
\bibinfo{author}{Wegner, P.~J.} \emph{et~al.}
\newblock \bibinfo{title}{{NIF} final optics system: frequency conversion and
  beam conditioning}.
\newblock \bibinfo{pages}{180}, \doiprefix\url{10.1117/12.538481}
  (\bibinfo{address}{San Jose, Ca}, \bibinfo{year}{2004}).

\bibitem{zhang_recent_2020}
\bibinfo{author}{Zhang, S.} \emph{et~al.}
\newblock \bibinfo{journal}{\bibinfo{title}{Recent advances in nonlinear optics
  for bio-imaging applications}}.
\newblock {\emph{\JournalTitle{Opto-Electronic Advances}}}
  \textbf{\bibinfo{volume}{3}}, \bibinfo{pages}{200003--200003},
  \doiprefix\url{10.29026/oea.2020.200003} (\bibinfo{year}{2020}).

\bibitem{dudley2024nonlinear}
\bibinfo{author}{Dudley, J.~M.}, \bibinfo{author}{Peacock, A.~C.},
  \bibinfo{author}{Stiller, B.} \& \bibinfo{author}{Tissoni, G.}
\newblock \bibinfo{title}{Nonlinear optics and its applications 2024}.
\newblock In \emph{\bibinfo{booktitle}{Proc. of SPIE Vol}}, vol.
  \bibinfo{volume}{13004}, \bibinfo{pages}{1300401--1} (\bibinfo{year}{2024}).

\bibitem{kwiat_new_1995}
\bibinfo{author}{Kwiat, P.~G.} \emph{et~al.}
\newblock \bibinfo{journal}{\bibinfo{title}{New {High}-{Intensity} {Source} of
  {Polarization}-{Entangled} {Photon} {Pairs}}}.
\newblock {\emph{\JournalTitle{Physical Review Letters}}}
  \textbf{\bibinfo{volume}{75}}, \bibinfo{pages}{4337--4341},
  \doiprefix\url{10.1103/PhysRevLett.75.4337} (\bibinfo{year}{1995}).

\bibitem{rubin_theory_1994}
\bibinfo{author}{Rubin, M.~H.}, \bibinfo{author}{Klyshko, D.~N.},
  \bibinfo{author}{Shih, Y.~H.} \& \bibinfo{author}{Sergienko, A.~V.}
\newblock \bibinfo{journal}{\bibinfo{title}{Theory of two-photon entanglement
  in type-{II} optical parametric down-conversion}}.
\newblock {\emph{\JournalTitle{Physical Review A}}}
  \textbf{\bibinfo{volume}{50}}, \bibinfo{pages}{5122--5133},
  \doiprefix\url{10.1103/PhysRevA.50.5122} (\bibinfo{year}{1994}).

\bibitem{valencia_large-scale_2025}
\bibinfo{author}{Valencia, N.~H.} \emph{et~al.}
\newblock \bibinfo{journal}{\bibinfo{title}{A large-scale reconfigurable
  multiplexed quantum photonic network}}.
\newblock {\emph{\JournalTitle{Nature Photonics}}}
  \doiprefix\url{10.1038/s41566-025-01806-x} (\bibinfo{year}{2025}).

\bibitem{lu_counter-propagating_2025}
\bibinfo{author}{Lu, Z.} \emph{et~al.}
\newblock \bibinfo{journal}{\bibinfo{title}{Counter-propagating entangled
  photon pairs from monolayer {GaSe}}}.
\newblock {\emph{\JournalTitle{Nature Communications}}}
  \textbf{\bibinfo{volume}{16}}, \bibinfo{pages}{9616},
  \doiprefix\url{10.1038/s41467-025-64620-7} (\bibinfo{year}{2025}).

\bibitem{chakraborty_towards_2025}
\bibinfo{author}{Chakraborty, T.} \emph{et~al.}
\newblock \bibinfo{journal}{\bibinfo{title}{Towards a spectrally multiplexed
  quantum repeater}}.
\newblock {\emph{\JournalTitle{npj Quantum Information}}}
  \textbf{\bibinfo{volume}{11}}, \bibinfo{pages}{3},
  \doiprefix\url{10.1038/s41534-024-00946-2} (\bibinfo{year}{2025}).

\bibitem{lyu_tunable_2025}
\bibinfo{author}{Lyu, X.} \emph{et~al.}
\newblock \bibinfo{journal}{\bibinfo{title}{A tunable entangled photon-pair
  source based on a {Van} der {Waals} insulator}}.
\newblock {\emph{\JournalTitle{Nature Communications}}}
  \textbf{\bibinfo{volume}{16}}, \bibinfo{pages}{1899},
  \doiprefix\url{10.1038/s41467-025-56436-2} (\bibinfo{year}{2025}).

\bibitem{hopfield_theory_1958}
\bibinfo{author}{Hopfield, J.~J.}
\newblock \bibinfo{journal}{\bibinfo{title}{Theory of the {Contribution} of
  {Excitons} to the {Complex} {Dielectric} {Constant} of {Crystals}}}.
\newblock {\emph{\JournalTitle{Physical Review}}}
  \textbf{\bibinfo{volume}{112}}, \bibinfo{pages}{1555--1567},
  \doiprefix\url{10.1103/PhysRev.112.1555} (\bibinfo{year}{1958}).

\bibitem{basov_polariton_2020}
\bibinfo{author}{Basov, D.~N.}, \bibinfo{author}{Asenjo-Garcia, A.},
  \bibinfo{author}{Schuck, P.~J.}, \bibinfo{author}{Zhu, X.} \&
  \bibinfo{author}{Rubio, A.}
\newblock \bibinfo{journal}{\bibinfo{title}{Polariton panorama}}.
\newblock {\emph{\JournalTitle{Nanophotonics}}} \textbf{\bibinfo{volume}{10}},
  \bibinfo{pages}{549--577}, \doiprefix\url{10.1515/nanoph-2020-0449}
  (\bibinfo{year}{2020}).

\bibitem{klyshko_scattering_nodate}
\bibinfo{author}{Klyshko, D.~N.}
\newblock \bibinfo{journal}{\bibinfo{title}{Scattering of light in a medium
  with nonlinear polarizability}}.
\newblock {\emph{\JournalTitle{JETP}}} \textbf{\bibinfo{volume}{28}}
  (\bibinfo{year}{1968}).

\bibitem{krebs_x-ray_2025}
\bibinfo{author}{Krebs, D.} \emph{et~al.}
\newblock \bibinfo{journal}{\bibinfo{title}{X-ray parametric down-conversion
  reveals {EUV}-polariton}}.
\newblock {\emph{\JournalTitle{Nature Communications}}}
  \textbf{\bibinfo{volume}{16}}, \doiprefix\url{10.1038/s41467-025-60845-8}
  (\bibinfo{year}{2025}).
\newblock \bibinfo{note}{Publisher: Springer Science and Business Media LLC}.

\bibitem{coffinet_coherent_1969}
\bibinfo{author}{Coffinet, J.~P.} \& \bibinfo{author}{De~Martini, F.}
\newblock \bibinfo{journal}{\bibinfo{title}{Coherent {Excitation} of
  {Polaritons} in {Gallium} {Phosphide}}}.
\newblock {\emph{\JournalTitle{Physical Review Letters}}}
  \textbf{\bibinfo{volume}{22}}, \bibinfo{pages}{60--64},
  \doiprefix\url{10.1103/PhysRevLett.22.60} (\bibinfo{year}{1969}).

\bibitem{kulevsky_light_1975}
\bibinfo{author}{Kulevsky, L.~A.}, \bibinfo{author}{Polivanov, Y.~N.} \&
  \bibinfo{author}{Poluektov, S.~N.}
\newblock \bibinfo{journal}{\bibinfo{title}{Light scattering by polaritons in
  {LiIO}$_{\textrm{3}}$}}.
\newblock {\emph{\JournalTitle{Journal of Raman Spectroscopy}}}
  \textbf{\bibinfo{volume}{3}}, \bibinfo{pages}{239--254},
  \doiprefix\url{10.1002/jrs.1250030213} (\bibinfo{year}{1975}).

\bibitem{aktsipetrov_frequency-angle_nodate}
\bibinfo{author}{Aktsipetrov, A.}, \bibinfo{author}{Ivanov, V.} \&
  \bibinfo{author}{Panin, A.}
\newblock \bibinfo{journal}{\bibinfo{title}{Frequency-angle spectrum of light
  scattering by polaritons and interference of susceptibilities of different
  orders}}.
\newblock {\emph{\JournalTitle{Zh. Eksp. Teor. Fiz}}}
  \textbf{\bibinfo{volume}{78}}, \bibinfo{pages}{2309--2315}
  (\bibinfo{year}{1980}).

\bibitem{chekhova_study_1993}
\bibinfo{author}{Chekhova, M.~V.} \& \bibinfo{author}{Penin, A.~N.}
\newblock \bibinfo{journal}{\bibinfo{title}{Study of second‐order excitations
  in $\alpha$‐iodic acid crystal by means of polariton \textit{k}
  ‐spectroscopy}}.
\newblock {\emph{\JournalTitle{Journal of Raman Spectroscopy}}}
  \textbf{\bibinfo{volume}{24}}, \bibinfo{pages}{581--584},
  \doiprefix\url{10.1002/jrs.1250240904} (\bibinfo{year}{1993}).

\bibitem{huotari_large-solid-angle_2017}
\bibinfo{author}{Huotari, S.} \emph{et~al.}
\newblock \bibinfo{journal}{\bibinfo{title}{A large-solid-angle {X}-ray {Raman}
  scattering spectrometer at {ID20} of the {European} {Synchrotron} {Radiation}
  {Facility}}}.
\newblock {\emph{\JournalTitle{Journal of Synchrotron Radiation}}}
  \textbf{\bibinfo{volume}{24}}, \bibinfo{pages}{521--530},
  \doiprefix\url{10.1107/S1600577516020579} (\bibinfo{year}{2017}).

\bibitem{henry_raman_1965}
\bibinfo{author}{Henry, C.~H.} \& \bibinfo{author}{Hopfield, J.~J.}
\newblock \bibinfo{journal}{\bibinfo{title}{Raman {Scattering} by
  {Polaritons}}}.
\newblock {\emph{\JournalTitle{Physical Review Letters}}}
  \textbf{\bibinfo{volume}{15}}, \bibinfo{pages}{964--966},
  \doiprefix\url{10.1103/PhysRevLett.15.964} (\bibinfo{year}{1965}).

\bibitem{sahle_planning_2015}
\bibinfo{author}{Sahle, C.~J.} \emph{et~al.}
\newblock \bibinfo{journal}{\bibinfo{title}{Planning, performing and analyzing
  {X}-ray {Raman} scattering experiments}}.
\newblock {\emph{\JournalTitle{Journal of Synchrotron Radiation}}}
  \textbf{\bibinfo{volume}{22}}, \bibinfo{pages}{400--409},
  \doiprefix\url{10.1107/S1600577514027581} (\bibinfo{year}{2015}).

\bibitem{galambosi_symmetry_2007}
\bibinfo{author}{Galambosi, S.}, \bibinfo{author}{Soininen, J.~A.},
  \bibinfo{author}{Nygård, K.}, \bibinfo{author}{Huotari, S.} \&
  \bibinfo{author}{Hämäläinen, K.}
\newblock \bibinfo{journal}{\bibinfo{title}{Symmetry of the 1 s core exciton in
  diamond studied using x-ray {Raman} scattering}}.
\newblock {\emph{\JournalTitle{Physical Review B}}}
  \textbf{\bibinfo{volume}{76}}, \bibinfo{pages}{195112},
  \doiprefix\url{10.1103/PhysRevB.76.195112} (\bibinfo{year}{2007}).

\bibitem{morar_observation_1985}
\bibinfo{author}{Morar, J.~F.}, \bibinfo{author}{Himpsel, F.~J.},
  \bibinfo{author}{Hollinger, G.}, \bibinfo{author}{Hughes, G.} \&
  \bibinfo{author}{Jordan, J.~L.}
\newblock \bibinfo{journal}{\bibinfo{title}{Observation of a {C}- 1 s {Core}
  {Exciton} in {Diamond}}}.
\newblock {\emph{\JournalTitle{Physical Review Letters}}}
  \textbf{\bibinfo{volume}{54}}, \bibinfo{pages}{1960--1963},
  \doiprefix\url{10.1103/PhysRevLett.54.1960} (\bibinfo{year}{1985}).

\bibitem{fano_effects_1961}
\bibinfo{author}{Fano, U.}
\newblock \bibinfo{journal}{\bibinfo{title}{Effects of {Configuration}
  {Interaction} on {Intensities} and {Phase} {Shifts}}}.
\newblock {\emph{\JournalTitle{Physical Review}}}
  \textbf{\bibinfo{volume}{124}}, \bibinfo{pages}{1866--1878},
  \doiprefix\url{10.1103/PhysRev.124.1866} (\bibinfo{year}{1961}).

\bibitem{weinhardt_resonant_2009}
\bibinfo{author}{Weinhardt, L.} \emph{et~al.}
\newblock \bibinfo{journal}{\bibinfo{title}{Resonant inelastic soft x-ray
  scattering of {CdS}: {A} two-dimensional electronic structure map approach}}.
\newblock {\emph{\JournalTitle{Physical Review B}}}
  \textbf{\bibinfo{volume}{79}}, \bibinfo{pages}{165305},
  \doiprefix\url{10.1103/PhysRevB.79.165305} (\bibinfo{year}{2009}).

\bibitem{doi:10.1366/0003702934067694}
\bibinfo{author}{Noda, I.}
\newblock \bibinfo{journal}{\bibinfo{title}{Generalized two-dimensional
  correlation method applicable to infrared, raman, and other types of
  spectroscopy}}.
\newblock {\emph{\JournalTitle{Applied Spectroscopy}}}
  \textbf{\bibinfo{volume}{47}}, \bibinfo{pages}{1329--1336},
  \doiprefix\url{10.1366/0003702934067694} (\bibinfo{year}{1993}).
\newblock \eprint{https://doi.org/10.1366/0003702934067694}.

\bibitem{huang_lattice_1951}
\bibinfo{author}{Huang, K.}
\newblock \bibinfo{journal}{\bibinfo{title}{Lattice {Vibrations} and {Optical}
  {Waves} in {Ionic} {Crystals}}}.
\newblock {\emph{\JournalTitle{Nature}}} \textbf{\bibinfo{volume}{167}},
  \bibinfo{pages}{779--780}, \doiprefix\url{10.1038/167779b0}
  (\bibinfo{year}{1951}).

\bibitem{mandal_theoretical_2023}
\bibinfo{author}{Mandal, A.} \emph{et~al.}
\newblock \bibinfo{journal}{\bibinfo{title}{Theoretical {Advances} in
  {Polariton} {Chemistry} and {Molecular} {Cavity} {Quantum}
  {Electrodynamics}}}.
\newblock {\emph{\JournalTitle{Chemical Reviews}}}
  \textbf{\bibinfo{volume}{123}}, \bibinfo{pages}{9786--9879},
  \doiprefix\url{10.1021/acs.chemrev.2c00855} (\bibinfo{year}{2023}).

\bibitem{torma_strong_2015}
\bibinfo{author}{Törmä, P.} \& \bibinfo{author}{Barnes, W.~L.}
\newblock \bibinfo{journal}{\bibinfo{title}{Strong coupling between surface
  plasmon polaritons and emitters: a review}}.
\newblock {\emph{\JournalTitle{Reports on Progress in Physics}}}
  \textbf{\bibinfo{volume}{78}}, \bibinfo{pages}{013901},
  \doiprefix\url{10.1088/0034-4885/78/1/013901} (\bibinfo{year}{2015}).

\bibitem{ebbesen_hybrid_2016}
\bibinfo{author}{Ebbesen, T.~W.}
\newblock \bibinfo{journal}{\bibinfo{title}{Hybrid {Light}–{Matter} {States}
  in a {Molecular} and {Material} {Science} {Perspective}}}.
\newblock {\emph{\JournalTitle{Accounts of Chemical Research}}}
  \textbf{\bibinfo{volume}{49}}, \bibinfo{pages}{2403--2412},
  \doiprefix\url{10.1021/acs.accounts.6b00295} (\bibinfo{year}{2016}).

\bibitem{toffoletti_coherent_2025}
\bibinfo{author}{Toffoletti, F.} \& \bibinfo{author}{Collini, E.}
\newblock \bibinfo{journal}{\bibinfo{title}{Coherent phenomena in
  exciton–polariton systems}}.
\newblock {\emph{\JournalTitle{Journal of Physics: Materials}}}
  \textbf{\bibinfo{volume}{8}}, \bibinfo{pages}{022002},
  \doiprefix\url{10.1088/2515-7639/adcbd6} (\bibinfo{year}{2025}).

\bibitem{baranov_novel_2018}
\bibinfo{author}{Baranov, D.~G.}, \bibinfo{author}{Wersäll, M.},
  \bibinfo{author}{Cuadra, J.}, \bibinfo{author}{Antosiewicz, T.~J.} \&
  \bibinfo{author}{Shegai, T.}
\newblock \bibinfo{journal}{\bibinfo{title}{Novel {Nanostructures} and
  {Materials} for {Strong} {Light}–{Matter} {Interactions}}}.
\newblock {\emph{\JournalTitle{ACS Photonics}}} \textbf{\bibinfo{volume}{5}},
  \bibinfo{pages}{24--42}, \doiprefix\url{10.1021/acsphotonics.7b00674}
  (\bibinfo{year}{2018}).

\bibitem{blaha_beyond_2022}
\bibinfo{author}{Blaha, M.}, \bibinfo{author}{Johnson, A.},
  \bibinfo{author}{Rauschenbeutel, A.} \& \bibinfo{author}{Volz, J.}
\newblock \bibinfo{journal}{\bibinfo{title}{Beyond the {Tavis}-{Cummings}
  model: {Revisiting} cavity {QED} with ensembles of quantum emitters}}.
\newblock {\emph{\JournalTitle{Physical Review A}}}
  \textbf{\bibinfo{volume}{105}}, \bibinfo{pages}{013719},
  \doiprefix\url{10.1103/PhysRevA.105.013719} (\bibinfo{year}{2022}).

\bibitem{schulke_electron_2007}
\bibinfo{author}{Schülke, W.}
\newblock \emph{\bibinfo{title}{Electron dynamics by inelastic {X}-ray
  scattering}}.
\newblock No.~\bibinfo{number}{7} in \bibinfo{series}{Oxford series on
  synchrotron radiation} (\bibinfo{publisher}{Oxford University Press},
  \bibinfo{address}{Oxford ; New York}, \bibinfo{year}{2007}).
\newblock \bibinfo{note}{OCLC: ocm85862430}.

\bibitem{boemer_towards_2021}
\bibinfo{author}{Boemer, C.} \emph{et~al.}
\newblock \bibinfo{journal}{\bibinfo{title}{Towards novel probes for valence
  charges \textit{via} {X}-ray optical wave mixing}}.
\newblock {\emph{\JournalTitle{Faraday Discussions}}}
  \textbf{\bibinfo{volume}{228}}, \bibinfo{pages}{451--469},
  \doiprefix\url{10.1039/D0FD00130A} (\bibinfo{year}{2021}).

\bibitem{PhysRevB.4.3610}
\bibinfo{author}{Painter, G.~S.}, \bibinfo{author}{Ellis, D.~E.} \&
  \bibinfo{author}{Lubinsky, A.~R.}
\newblock \bibinfo{journal}{\bibinfo{title}{Ab initio calculation of the
  electronic structure and optical properties of diamond using the discrete
  variational method}}.
\newblock {\emph{\JournalTitle{Phys. Rev. B}}} \textbf{\bibinfo{volume}{4}},
  \bibinfo{pages}{3610--3622}, \doiprefix\url{10.1103/PhysRevB.4.3610}
  (\bibinfo{year}{1971}).

\bibitem{tamasaku_determining_2009}
\bibinfo{author}{Tamasaku, K.}, \bibinfo{author}{Sawada, K.} \&
  \bibinfo{author}{Ishikawa, T.}
\newblock \bibinfo{journal}{\bibinfo{title}{Determining {X}-{Ray} {Nonlinear}
  {Susceptibility} of {Diamond} by the {Optical} {Fano} {Effect}}}.
\newblock {\emph{\JournalTitle{Physical Review Letters}}}
  \textbf{\bibinfo{volume}{103}}, \bibinfo{pages}{254801},
  \doiprefix\url{10.1103/PhysRevLett.103.254801} (\bibinfo{year}{2009}).

\bibitem{garcia-vidal_manipulating_2021}
\bibinfo{author}{Garcia-Vidal, F.~J.}, \bibinfo{author}{Ciuti, C.} \&
  \bibinfo{author}{Ebbesen, T.~W.}
\newblock \bibinfo{journal}{\bibinfo{title}{Manipulating matter by strong
  coupling to vacuum fields}}.
\newblock {\emph{\JournalTitle{Science}}} \textbf{\bibinfo{volume}{373}},
  \bibinfo{pages}{eabd0336}, \doiprefix\url{10.1126/science.abd0336}
  (\bibinfo{year}{2021}).

\bibitem{sanchez-barquilla_theoretical_2022}
\bibinfo{author}{Sánchez-Barquilla, M.},
  \bibinfo{author}{Fernández-Domínguez, A.~I.}, \bibinfo{author}{Feist, J.}
  \& \bibinfo{author}{García-Vidal, F.~J.}
\newblock \bibinfo{journal}{\bibinfo{title}{A {Theoretical} {Perspective} on
  {Molecular} {Polaritonics}}}.
\newblock {\emph{\JournalTitle{ACS Photonics}}} \textbf{\bibinfo{volume}{9}},
  \bibinfo{pages}{1830--1841}, \doiprefix\url{10.1021/acsphotonics.2c00048}
  (\bibinfo{year}{2022}).

\bibitem{li_molecular_2022}
\bibinfo{author}{Li, T.~E.}, \bibinfo{author}{Cui, B.},
  \bibinfo{author}{Subotnik, J.~E.} \& \bibinfo{author}{Nitzan, A.}
\newblock \bibinfo{journal}{\bibinfo{title}{Molecular {Polaritonics}:
  {Chemical} {Dynamics} {Under} {Strong} {Light}–{Matter} {Coupling}}}.
\newblock {\emph{\JournalTitle{Annual Review of Physical Chemistry}}}
  \textbf{\bibinfo{volume}{73}}, \bibinfo{pages}{43--71},
  \doiprefix\url{10.1146/annurev-physchem-090519-042621}
  (\bibinfo{year}{2022}).

\bibitem{nithianandam_synchrotron_1993}
\bibinfo{author}{Nithianandam, J.} \& \bibinfo{author}{Rife, J.~C.}
\newblock \bibinfo{journal}{\bibinfo{title}{Synchrotron x-ray optical
  properties of natural diamond}}.
\newblock {\emph{\JournalTitle{Physical Review B}}}
  \textbf{\bibinfo{volume}{47}}, \bibinfo{pages}{3517--3521},
  \doiprefix\url{10.1103/PhysRevB.47.3517} (\bibinfo{year}{1993}).

\bibitem{blum_ab_2009}
\bibinfo{author}{Blum, V.} \emph{et~al.}
\newblock \bibinfo{journal}{\bibinfo{title}{Ab initio molecular simulations
  with numeric atom-centered orbitals}}.
\newblock {\emph{\JournalTitle{Computer Physics Communications}}}
  \textbf{\bibinfo{volume}{180}}, \bibinfo{pages}{2175--2196},
  \doiprefix\url{10.1016/j.cpc.2009.06.022} (\bibinfo{year}{2009}).

\bibitem{ren_resolution--identity_2012}
\bibinfo{author}{Ren, X.} \emph{et~al.}
\newblock \bibinfo{journal}{\bibinfo{title}{Resolution-of-identity approach to
  {Hartree}–{Fock}, hybrid density functionals, {RPA}, {MP2} and
  \textit{{GW}} with numeric atom-centered orbital basis functions}}.
\newblock {\emph{\JournalTitle{New Journal of Physics}}}
  \textbf{\bibinfo{volume}{14}}, \bibinfo{pages}{053020},
  \doiprefix\url{10.1088/1367-2630/14/5/053020} (\bibinfo{year}{2012}).

\bibitem{heyd_hybrid_2003}
\bibinfo{author}{Heyd, J.}, \bibinfo{author}{Scuseria, G.~E.} \&
  \bibinfo{author}{Ernzerhof, M.}
\newblock \bibinfo{journal}{\bibinfo{title}{Hybrid functionals based on a
  screened {Coulomb} potential}}.
\newblock {\emph{\JournalTitle{The Journal of Chemical Physics}}}
  \textbf{\bibinfo{volume}{118}}, \bibinfo{pages}{8207--8215},
  \doiprefix\url{10.1063/1.1564060} (\bibinfo{year}{2003}).

\bibitem{schulke_interband_1988}
\bibinfo{author}{Schülke, W.}, \bibinfo{author}{Bonse, U.},
  \bibinfo{author}{Nagasawa, H.}, \bibinfo{author}{Kaprolat, A.} \&
  \bibinfo{author}{Berthold, A.}
\newblock \bibinfo{journal}{\bibinfo{title}{Interband transitions and core
  excitation in highly oriented pyrolytic graphite studied by inelastic
  synchrotron x-ray scattering: {Band}-structure information}}.
\newblock {\emph{\JournalTitle{Physical Review B}}}
  \textbf{\bibinfo{volume}{38}}, \bibinfo{pages}{2112--2123},
  \doiprefix\url{10.1103/PhysRevB.38.2112} (\bibinfo{year}{1988}).

\bibitem{mcdougall_near_2014}
\bibinfo{author}{McDougall, N.~L.}, \bibinfo{author}{Nicholls, R.~J.},
  \bibinfo{author}{Partridge, J.~G.} \& \bibinfo{author}{McCulloch, D.~G.}
\newblock \bibinfo{journal}{\bibinfo{title}{The {Near} {Edge} {Structure} of
  {Hexagonal} {Boron} {Nitride}}}.
\newblock {\emph{\JournalTitle{Microscopy and Microanalysis}}}
  \textbf{\bibinfo{volume}{20}}, \bibinfo{pages}{1053--1059},
  \doiprefix\url{10.1017/S1431927614000737} (\bibinfo{year}{2014}).

\bibitem{soufli_reflectance_1997}
\bibinfo{author}{Soufli, R.} \& \bibinfo{author}{Gullikson, E.~M.}
\newblock \bibinfo{journal}{\bibinfo{title}{Reflectance measurements on clean
  surfaces for the determination of optical constants of silicon in the extreme
  ultraviolet–soft-x-ray region}}.
\newblock {\emph{\JournalTitle{Applied Optics}}} \textbf{\bibinfo{volume}{36}},
  \bibinfo{pages}{5499}, \doiprefix\url{10.1364/AO.36.005499}
  (\bibinfo{year}{1997}).

\bibitem{dorney_actinic_2024}
\bibinfo{author}{Dorney, K.~M.} \emph{et~al.}
\newblock \bibinfo{journal}{\bibinfo{title}{Actinic inspection of the extreme
  ultraviolet optical parameters of lithographic materials enabled by a
  table-top, coherent extreme ultraviolet source}}.
\newblock {\emph{\JournalTitle{Journal of Micro/Nanopatterning, Materials, and
  Metrology}}} \textbf{\bibinfo{volume}{23}},
  \doiprefix\url{10.1117/1.JMM.23.4.041406} (\bibinfo{year}{2024}).

\bibitem{tallents_lithography_2010}
\bibinfo{author}{Tallents, G.}, \bibinfo{author}{Wagenaars, E.} \&
  \bibinfo{author}{Pert, G.}
\newblock \bibinfo{journal}{\bibinfo{title}{Lithography at {EUV} wavelengths}}.
\newblock {\emph{\JournalTitle{Nature Photonics}}}
  \textbf{\bibinfo{volume}{4}}, \bibinfo{pages}{809--811},
  \doiprefix\url{10.1038/nphoton.2010.277} (\bibinfo{year}{2010}).

\bibitem{wagner_lithography_2010}
\bibinfo{author}{Wagner, C.} \& \bibinfo{author}{Harned, N.}
\newblock \bibinfo{journal}{\bibinfo{title}{Lithography gets extreme}}.
\newblock {\emph{\JournalTitle{Nature Photonics}}}
  \textbf{\bibinfo{volume}{4}}, \bibinfo{pages}{24--26},
  \doiprefix\url{10.1038/nphoton.2009.251} (\bibinfo{year}{2010}).

\bibitem{ciesielski_determination_2022}
\bibinfo{author}{Ciesielski, R.} \emph{et~al.}
\newblock \bibinfo{journal}{\bibinfo{title}{Determination of optical constants
  of thin films in the {EUV}}}.
\newblock {\emph{\JournalTitle{Applied Optics}}} \textbf{\bibinfo{volume}{61}},
  \bibinfo{pages}{2060}, \doiprefix\url{10.1364/AO.447152}
  (\bibinfo{year}{2022}).

\bibitem{wood_improved_2014}
\bibinfo{author}{Tarrio, C.} \emph{et~al.}
\newblock \bibinfo{title}{Improved measurement capabilities at the {NIST} {EUV}
  reflectometry facility}.
\newblock \bibinfo{pages}{90481I}, \doiprefix\url{10.1117/12.2046290}
  (\bibinfo{address}{San Jose, California, United States},
  \bibinfo{year}{2014}).

\bibitem{sokolov_at-wavelength_2016}
\bibinfo{author}{Sokolov, A.} \emph{et~al.}
\newblock \bibinfo{journal}{\bibinfo{title}{At-wavelength metrology facility
  for soft {X}-ray reflection optics}}.
\newblock {\emph{\JournalTitle{Review of Scientific Instruments}}}
  \textbf{\bibinfo{volume}{87}}, \bibinfo{pages}{052005},
  \doiprefix\url{10.1063/1.4950731} (\bibinfo{year}{2016}).

\bibitem{behringer_characterization_2015}
\bibinfo{author}{Laubis, C.}, \bibinfo{author}{Haase, A.},
  \bibinfo{author}{Soltwisch, V.} \& \bibinfo{author}{Scholze, F.}
\newblock \bibinfo{title}{Characterization of optical material parameters for
  {EUV} {Lithography} applications at {PTB}}.
\newblock \bibinfo{pages}{96610W}, \doiprefix\url{10.1117/12.2195009}
  (\bibinfo{address}{Eindhoven, Netherlands}, \bibinfo{year}{2015}).

\end{thebibliography}



\section*{Acknowledgements}
We acknowledge the European Synchrotron Radiation Facility (ESRF) for provision of synchrotron radiation facilities  under proposal number HC-5736 and we would like to thank F. Gerbon for assistance and support in using beamline ID20.
The authors gratefully acknowledge support from T. Liedke and A. Barinskaya during the beamtime.

\section*{Author contributions statement}
F.K., D.K., C.J.S., B.D. and C.B. conducted the experiments, 
D.K. developed the theoretical description and F.K. performed modelling and data analysis 
with help by X.B. and A.N. and supervision by D.K. and C.B..  
F.K., D.K. and C.B. wrote the manuscript with contributions from all authors. 
The co-first authorship order was determined alphabetically based on surnames. Both F.K. and D.K. contributed equally and have the right to list their name first in their CV.

\section*{Additional information}

\textbf{Accession codes} The experimental data used for this study have been deposited in the Zenodo repository under accession code \href{http://doi.org//10.5281/zenodo.17910813}{10.5281/zenodo.17910813}.\\
\textbf{Competing interests} The authors declare no competing interests. 
\end{document}